\documentclass[sn-mathphys,Numbered]{sn-jnl}

\usepackage{graphicx}%
\usepackage{multirow}%
\usepackage{amsmath,amssymb,amsfonts}%
\usepackage{amsthm}%
\usepackage{mathrsfs}%
\usepackage[title]{appendix}%
\usepackage{xcolor}%
\usepackage{textcomp}%
\usepackage{manyfoot}%
\usepackage{booktabs}%
\usepackage{algorithm}%
\usepackage{algorithmicx}%
\usepackage{algpseudocode}%
\usepackage{listings}%
\usepackage{booktabs}
\usepackage[utf8]{inputenc} 

\theoremstyle{thmstyleone}%
\theoremstyle{thmstyletwo}%

\theoremstyle{thmstylethree}%

\begin{document}

\title[Article Title]{Science as Exploration in a Knowledge Landscape: Tracing Hotspots or Seeking Opportunity?}


\author[1,2]{\fnm{Feifan} \sur{Liu}}\email{liufeifan@dlut.edu.cn}

\author[1,2]{\fnm{Shuang} \sur{Zhang}}\email{shuang94@mail.dlut.edu.cn}

\author*[1,2,3,4]{\fnm{Haoxiang} \sur{Xia}}\email{hxxia@dlut.edu.cn}

\affil*[1]{\orgdiv{Institute of Systems Engineering}, \orgname{Dalian University of Technology}, \orgaddress{\street{No.2 Linggong Road}, \city{Dalian}, \postcode{116024}, \state{Liaoning}, \country{China}}}

\affil[2]{\orgdiv{Center for Big Data and Intelligent Decision-Making}, \orgname{Dalian University of Technology}, \orgaddress{\street{Linggong Road}, \city{Dalian}, \postcode{116024}, \state{Liaoning}, \country{China}}}
\affil[3]{\orgdiv{Institute for Advanced Intelligence}, \orgname{Dalian University of Technology}, \orgaddress{\street{Linggong Road}, \city{Dalian}, \postcode{116024}, \state{Liaoning}, \country{China}}}
\affil[4]{\orgdiv{Key Laboratory of Social Computing and Cognitive Intelligence}, \orgname{Ministry of Education of China}, \orgaddress{\street{Linggong Road}, \city{Dalian}, \postcode{116024}, \state{Liaoning}, \country{China}}}



\abstract{The selection of research topics by scientists can be viewed as an exploration process conducted by individuals with cognitive limitations traversing a complex cognitive landscape influenced by both individual and social factors. While existing theoretical investigations have provided valuable insights, the intricate and multifaceted nature of modern science hinders the implementation of empirical experiments. This study leverages advancements in deep learning techniques to investigate the patterns and dynamic mechanisms of topic-transition among scientists. By constructing the knowledge space across 6 large-scale disciplines, we depict the trajectories of scientists' topic transitions within this space, measuring the flow and distance of research regions across different sub-spaces. Our findings reveal a predominantly conservative pattern of topic transition at the individual level, with scientists primarily exploring local knowledge spaces. Furthermore, simulation modeling analysis identifies research intensity, driven by the concentration of scientists within a specific region, as the key facilitator of topic transition. Conversely, the knowledge distance between fields serves as a significant barrier to exploration. Notably, despite potential opportunities for breakthrough discoveries at the intersection of subfields, empirical evidence suggests that these opportunities do not exert a strong pull on scientists, leading them to favor familiar research areas. Our study provides valuable insights into the exploration dynamics of scientific knowledge production, highlighting the influence of individual cognition, social factors, and the intrinsic structure of the knowledge landscape itself. These findings offer a framework for understanding and potentially shaping the course of scientific progress.}

\keywords{Scientists’ exploration, Knowledge space, Topic-transition behavior, Gravity model, Radiation model}



\maketitle

\section{Introduction}\label{sec1}

Throughout their academic careers, scientists must confront a multitude of choices when it comes to selecting their research topics. These decisions wield a substantial influence over their academic productivity, impact, and overall career trajectory. Nobel laureate Chen Ning Yang shared a valuable insight during a symposium at the University of Chinese Academy of Sciences\cite{RN1847}. He emphasized that, particularly for emerging scientists, the decision to persist in a particular field may not directly dictate their career's level of achievement. However, the wise selection of research topics and research directions holds paramount significance. In his words, \begin{quote}
    \textit{Pursuing a direction that leads to an impasse can be a treacherous endeavor, as the deeper one delves, the more arduous it becomes to alter course. Diverting from an unproductive trajectory is no simple feat, making persistence in a barren direction a most regrettable choice.}
\end{quote}On a broader scale, the choices made by scientists in terms of topic selection and transition impact the development of the entire scientific ecosystem. Understanding the intricate motivations and multifaceted influences that guide scientists' decisions in the process of selecting research topics presents a substantial challenge in unraveling the behavioral patterns and internal mechanisms that underlie these choices.\par
Scientists' choices of topics can be illuminated as the persistent endeavors of cognitively constrained individuals within the intricate expanse of knowledge\cite{RN811}. This pursuit adheres to the principle of “no free lunch”. Owing to the inherent tension between accumulating academic accomplishments and fostering innovation, scientists grapple with the delicate task of balancing conventional and pioneering research fields\cite{RN1843}. Diverse strategies employed in the process of topic selection yield markedly distinct outcomes, impacting both personal development \cite{RN245, Yu2021, Huang2023}and scientific progress\cite{RN634}. Consequently, various levels of behavioral risk must be contemplated. To unravel these intrinsic conundrums, prior investigations have empirically validated and dissected the trade-offs scientists encounter during their exploration, focusing primarily on individual scientists' topic selection and their relationship with academic performance within their respective research fields\cite{RN665, RN1370}.\par
The exploration within the realm of knowledge reflects a complex interplay of scientists' decision-making behaviors. The selection of research topics is shaped by individual volition and concurrently influenced by the collective dynamics within the specific knowledge field. In contrast to the early days of modern scientific development, characterized by a limited number of scientists who primarily pursued research based on personal interests, contemporary scientific progress has witnessed a proliferation of participants and a diversification of topic matter\cite{RN483, RN1191}. This expansion inevitably renders the process of selecting research topics susceptible to the impact of social factors. As government entities, corporations, and diverse social organizations have increasingly assumed central roles in funding scientific research, the defining characteristics of the scientific establishment have become more pronounced. In this era of ‘big science’, scientists' choice of topics is not solely propelled by personal aspirations and inclinations. It is equally shaped by a spectrum of social behaviors such as following, learning, emulating, and conforming to prevailing trends.\par
Aligning research interests within scholarly groups has the potential to accelerate scientific outputs, increase scholarly impact, and improve access to scholarly resources. This, in turn, serves the advancement of individual scholarly careers. However, it is important to remain vigilant that the advancement of science depends on groundbreaking discoveries and trendsetting contributions. An overemphasis on conforming to popular trends and crowd-sourced research selection may lead to stagnation within the broader scientific research and innovation ecosystem\cite{RN1473}, potentially resulting in a scenario where resources are allocated without commensurate progress. \par
The central question is whether scientists should opt for popular research areas that attract widespread attention or explore an uncultivated territory of research fields. It concerns the patterns of behavior that scientists exhibit when moving between topics within or across the research field. Can these patterns be quantified and further explained by a simple mechanistic model of group behavior? A comprehensive understanding of these issues can shed light on the strategic choices and risk preferences of scientists, provide deep insights into the underlying mechanisms of scientific development, and serve as a valuable basis for the design of research management policies.\par
To gain a deeper understanding of knowledge spaces and scientists' exploratory behaviors within them, we draw inspiration from Geographic Information Systems (GIS) principles. The analysis of human mobility patterns in physical space has provided valuable insights\cite{RN314}. Recent advancements in machine learning, especially in representation learning algorithms, have opened up opportunities for measuring knowledge distance between research subfields and help us better quantify the intricate and abstract knowledge spaces of disciplines\cite{RN1845}, underpinning the empirical study of the collective mobility behavior of scientists.\par
Therefore, to bridge the gap in understanding scientists' topic selection and transition patterns at the population level, this study builds on the foundation of constructing a scientific knowledge space as a research field map, and attempts to integrate complex network analysis methods, machine learning algorithms, and geographic information analysis theories to understand the collective knowledge creation process in the scientific ecosystem. The main research contributions of this paper are as follows: \par
(1) Within the framework of constructing a knowledge space, scientists' papers are embedded in this space based on the topical distance. The knowledge space is partitioned into the grid and the voronoi diagram subfields, using both equidistant and equal-density approaches. Scientists’  trajectories are constituted of published papers and merge into OD flows that effectively encapsulate scientists' exploration patterns in the knowledge space. The analysis of these topic selection and transition trajectories, when rooted in the entire scientific field space, provides novel insights for quantifying scientists' topic-changing. Including activities such as online socializing, web searching, and gaming, all of which involve complex and abstract spaces, the methodological approach in this study can potentially be extended to quantify individual-level or population-level mobility in virtual spaces with fine granularity.\par
(2) When exploring the flow of scientists' publication trajectories across different subfields within the knowledge space, it is evident that the distance traveled by scientists as they move between topics follows a log-normal distribution. This observation is particularly pronounced in the context of voronoi diagram-based field partitioning. This broad, “heavy-tailed” distribution suggests that scientists' inter-field movement patterns, while predominantly characterized by short-range transits, also include occasional long-range transitions. It is noteworthy, however, that these patterns do not exhibit a “scale-free” behavior, underscoring that the majority of scientists tend to change their subfields with cautious, short-range transits.\par
(3) Intriguingly, the study reveals that the gravity model, which takes into account factors such as population size and the distances between starting and ending points, offers a more robust explanation and prediction of scientists' topic selection and transition within the knowledge space. In the quest to unravel the underlying mechanisms governing scientists' topic-transition patterns at the group level, this study introduces two distinct group exploration models: the distance-based “gravity” model and the opportunity-based “radiation” model. Our finding implies that the fundamental driving force behind scientists' topic selection and change is the research hotspots generated by the density of scientists in a given region. Conversely, the inhibiting factor is the knowledge distance between distinct fields. While research opportunities may exist at the intersection of subfields, this factor does not significantly influence scientists' decisions to change their research focus.\par
In Section 2, we describe the use of the dataset, the framework for constructing a knowledge space, the tessellated diagram types of spatial partitioning, the gravity model, the radiation model, and corresponding evaluation metrics. In Section 3, we use complex network and representation learning techniques to construct a knowledge space for physics using the APS dataset and identify paper positions. We then use the grid and voronoi diagram to delineate sub-field regions, capturing the population-level mobility of scientists in the knowledge spaces. To disclose the underlying mechanism of scientists' inter-field OD flow, we introduce the gravity model and the radiation model. Then we test and validate the explanatory and predictive capabilities of these models on the mobile patterns of scientists in the knowledge space. In Section 4, we discuss our findings with studies on human mobility patterns in real and virtual spaces and other related works. Finally, in Section 5, we summarize our main findings, highlight research limitations, and suggest future directions.
\section{Materials and methods}
\subsection{Dataset}
The major part of this paper focuses on the field of physics and utilizes the journal literature dataset provided by the American Physical Society (APS)\cite{RN622}. In exploring the topic-transition behavior patterns of scientists, more than 258,000 papers published in APS journals from 1985 to 2009 were used. Taking into account the impact of authors and the percentage of the number of papers, 13,720 scientists in the field of physics with more than or equal to 16 publications, involving 450,290 publication records, were eventually selected. Author and paper records were preprocessed and provided by Sinatra et al \cite{RN214}. The selection of scientists is based on the fact that although the number of scientists with 16 or more publications accounts for only 13.1\% (13,720/104,483) of the dataset of this study, the number of their papers accounts for 82.4\% (209,473/254,117). \par
Our findings have also been further extended to Computer Science, Chemistry, Biology, Social Science, and Multidisciplinary Science with Microsoft Academic Graph(MAG)\cite{Sinha2015}. Leveraging the comprehensive “fields of study” classification system provided by the Microsoft Academic Graph (MAG) \cite{wang2019a}, we extract a dataset encompassing 4,752,206 authors and 4,391,220 papers associated with the label “Computer Science”, spanning from 1948 to 2019. Subsequently, we focus on a subset of 180,339 highly productive scientists, each with a minimum of 10 published papers within the domain. The Chemistry dataset encompassed 9,568,741 authors and 6,916,260 papers labeled “Chemistry”, covering the period until 2019. We focus our analysis on 117,960 prolific scientists who had published at least 30 papers, totally involved with 4,048,890 papers. The Biology dataset, comprising 9,731,092 authors and 7,157,231 papers categorized as “Biology” in MAG, covered the same timeframe. We finally identify 164,871 highly active scientists, whose papers count greater than or equal to 30, and their 4,701,836 papers. The Social Science dataset consisted of 740,196 authors and 765,709 papers published in journals belonging to the SAGE publishing group, spanning the period from 1965 to 2019. Our analysis focuses on 19,105 scientists, whose number of published papers is larger than or equal to 10, and their 237,278 papers in this domain. Furthermore, we construct a multidisciplinary dataset encompassing scientific publications from five prominent journals representing diverse research areas: Nature, Science, Proceedings of the National Academy of Sciences, Nature Communications, and Science Advances. This dataset comprises 948,180 authors and 562,998 papers published between 1869 and 2019. We identify 22,842 scientists, who had published at least 10 papers, contributing to a collective body of 295,888 papers in this area.
\subsection{Construction of knowledge space}
In the context of the scientific innovation system, a crucial aspect of the collective behavior of scientists corresponds to their decisions and transitions in research directions within the epistemic landscape. The establishment of an accurate and valid knowledge space serves as the basis for determining the distance at which scientists' interests change. 
Given the stable characteristic of most physical subfields\cite{RN1172}, we construct a knowledge network of physics disciplines by utilizing the co-occurrence relationship between PACS-level classification codes and their co-occurrence frequency in each paper published in APS journals. This network consists of 874 secondary PACS codes as nodes and co-occurrence relationships between PACS codes as connected edges. Considering the elimination of the influence of the absolute difference in frequency between PACS codes, we further take the square root of the inverse of the joint probability of PACS code \textit{i} and PACS code \textit{j} appearing in a paper at the same time as the weight value \(w_{ij}\) of the network, and the calculation process is shown in Equation (1):\par
\begin{equation}
 w_{ij}=\frac{1}{\sqrt{(\frac{f_{ij}}{f_i} \cdot \frac{f_{ij}}{f_j})}}=\frac{\sqrt{(f_i f_j )}}{f_{ij}}\label{eq1}
\end{equation}
where the \(f_i\) and \(f_j\) are the cumulative edge frequencies in the network connected to node \textit{i} and node \textit{j}, respectively. The network's modularity, calculated at approximately 0.506 through a community detection algorithm\cite{RN655}, signifies the presence of distinct community structures within the field of physics. This implies that physics can be divided into several closely related subfields with relatively sparse interconnections between them. We then apply Node2Vec\cite{RN925} and the UMAP manifold learning algorithm\cite{RN1013} to create a knowledge map of physics.  \par
Furthermore, to eliminate the potential influence of choosing representation methods for our observed patterns in this study, we utilize Doc2Vec\cite{Mikolov2013}, a widely used document embedding technique, to extract high-dimensional features from the title and abstract of research papers belonging to the other five disciplines. This approach ensures consistency across different disciplines and minimizes bias introduced by specific representation learning methods. The constructed map represents the research field and benefits from representation learning to uncover knowledge structure and manifold learning for virtual spatial analysis. Overall, this approach facilitates embedding and visualizing the scientific landscape and offers a foundation for quantifying scientific research movements within the knowledge space.
\subsection{Tessellated models of space: grid and voronoi diagram}
To comprehensively analyze the topic selection and transition of scientists, the following step involves partitioning the knowledge space into distinct regions and identifying the “geographic units”. In real-world geographic spaces, people often adopt administrative districts as their fundamental research units. However, these pre-defined districts do not exist within the realm of knowledge spaces. Consequently, in this section, the knowledge space is divided into spatial regions based on the principles of “equal distance” and “equal density”, with subsequent comparison of scientists' behavioral patterns.
Tessellated models of space, including grid and voronoi diagrams, serve as potent tools for the representation and analysis of spatial arrangements\cite{RN1449}. They offer a unified research framework for comprehending the knowledge space. In this study, we employ those two distinct spatial region delineation approaches to understand the impact of the knowledge space delineation method on our research conclusions.\par
The grid diagram approach involves partitioning the entire knowledge field map into a series of grid regions, with each grid region spanning a 1° interval in knowledge space. This results in a total of 90 grid regions arranged in a 10x9 configuration, of which 73 available non-empty grid regions were associated with the specific research areas addressed in this study.\par
On the other hand, the voronoi partitioning approach utilizes the spatial distribution of high-frequency PACS codes within co-occurrence networks to define the knowledge space. Initially, we identify the top 10 high-frequency PACS codes within each subfield region and designate their centroid positions as the focal points in the voronoi diagram field. These 90 positions were instrumental in generating the boundaries of the voronoi diagram.\par
The main difference between these two methods is their spatial division approach. The grid diagram method divides space into uniform grid points, maintaining an isometric structure. On the other hand, the voronoi diagram, determined by the high-frequency PACS code, divides space based on isodensity, aligning with the heterogeneous distribution of the population. In this study, we will perform statistical analyses of scientists' group mobility origin-destination (OD) flows and use predictive modeling to analyze trajectory patterns under both tessellated modes of knowledge spatial region.
\subsection{Models of OD flow prediction: Gravity Model and Radiation Model}
The Gravity Model\cite{RN1565} and the Radiation Model\cite{RN643} are two prominent mathematical models employed in human mobility and migration studies. These models aim to elucidate the population-level patterns of movement between different locations. The Gravity Model is predominantly distance-based, while the Radiation Model additionally incorporates factors like competition for destinations and accessibility.\par
Specifically, the gravity model, inspired by Newton's gravitational formula, suggests that the flow of exploration by groups in different regions is directly proportional to the size of the regional group and inversely proportional to the square of the distance accessible between regions\cite{zipf46}. The model was also the firstly used in the field of geography to explain group migration. The mathematical expression of the general gravity model is shown in equation (2):\par
\begin{equation}
T_{ij}=\frac{(m_i^\alpha) (n_j^\beta)}{f(d_{ij}) }
\end{equation}
where \(T_{ij}\) denotes the flow of people between location \textit{i} and location \textit{j}, \(m_i\) and \(n_j\) denote the total population of location \textit{i} and location \textit{j}, respectively. \(d_{ij}\) denotes the distance between locations \textit{i} and \textit{j}. \(\alpha\) and \(\beta\) are adjustable exponential variables. \(f(d_{ij})\) is a damping function set according to different empirical data, such as a power-law function \(f(d_{ij} )= d_{ij}^\gamma \) or exponential function \(f(d_{ij})= e^{(\gamma \cdot d_{ij})}\). Depending on the constraints, gravity models can also be categorized into models under one-way and two-way constraints. This type of constrained model can more accurately estimate and predict total inter-regional flows by fixing the population from location \textit{i} to location \textit{j} (output model) or the number of people entering (attraction model).\par
Inspired by the opportunity model, Simini et al.\cite{RN643} propose a radiation model that more accurately predicts population movement. They claim that the radiation model not only predicts the average flow between two locations but also captures the variability of the flow compared to the gravity model. Specifically, the mathematical expression of the radiation model is given in equation (3):\par
\begin{equation}
\langle T_{ij}\rangle= \frac{T_i(m_in_j)}{(m_i+s_{ij})(m_i+n_j+s_{ij})}
\end{equation}
where \(\langle T_{ij}\rangle\) denotes the average population flow between location \textit{i} and location \textit{j} and \(T_i\equiv \sum_{(j\neq i)}T_{ij} \). Compared to the gravity model, an additional parameter \(s_{ij}\) has been introduced. This parameter represents the population (or employment opportunities) outside of locations \textit{i} and \textit{j} within a distance of \(d_{ij}\). It signifies the potential opportunities within the range from location \textit{i} to location \textit{j} that attract people to move.\par
The gravity model is a one-way constraint model that predetermines the population size at the origin while incorporating power-law and exponential damping functions to capture varying distance effects. In contrast, the radiation model is a parameter-free model, and we directly apply Eq. (3) for conducting simulation experiments.
\subsection{The evaluation metrics of the population-level human mobility model}
To quantify the performance of population-level models in this study, we then introduce a set of evaluation metrics. Human mobility model evaluation metrics are specifically designed to gauge the level of consistency between a model and actual human mobility data within spatial contexts. Beyond the common metrics such as R-squared, root mean square error, Spearman's correlation coefficient, and Pearson's correlation coefficient, the evaluation metrics for human mobility behavior models also encompass distinctive measures for assessing the convergence of human mobile activities\cite{RN1434}. \par
These measures include the Common Part of Commuters (\textit{CPC}), which quantifies the proportion of individuals with overlapping trajectories, the Common Part of Commuters' Distance (\textit{\(CPC_d\)}), which represents the fraction of overlapping distances traveled, and the Common Part of Links (\textit{CPL}), which indicates the extent of overlap in mobility paths. Detailed formulas for computing these three metrics can be found in equations (4-6):\par
\begin{equation}
    CPC(T,\widetilde{T})=\frac{\sum_{(i,j=1)}^n min(T_{ij},\widetilde{T}_{ij})}{N}=1-\frac{1}{2}\frac{\sum_{(i,j=1)}^n |T_{ij}-\widetilde{T}_{ij}|}{N}
\end{equation}

\begin{equation}
CPC_d (T,\widetilde{T})=\frac{\sum_{(k=1)}^\infty min(N_k,\widetilde{N}_k)}{N}
\end{equation}
\begin{equation}
    CPL(T,\widetilde{T})= \frac{2\sum_{(i,j=1)}^n 1_{(T_{ij}>0)} \cdot 1_{(\widetilde{T}_{ij}>0))}}{\sum_{(i,j=1)}^n 1_{(T_{ij}>0)}+\sum_{(i,j=1)}^n 1_{(\widetilde{T}_{ij}>0)}}
\end{equation}
Among the three formulas mentioned earlier, the symbols T and \(\widetilde{T}\) represent the actual flow and model-predicted flow values between locations \textit{i} and \textit{j}, respectively. \textit{N} refers to the overall population flow, while \(N_k\) denotes the number of individual movements occurring between distances in the range of 2k-2 to 2k. The variable \(1_x\) takes on a value of 1 when condition \textit{x} is met, and it is 0 otherwise. \par
These indicators evaluate the precision of the model's fitting or predictions, considering three essential factors: the population size, the knowledge distance, and the particular routes. These scores are instrumental in identifying the model's strengths and limitations, as well as its adaptability for a specific human movement context at the population level.
\section{Results}
\subsection{Knowledge space and trajectories in Physics}
Using the embedded PACS code co-occurrence network as a foundation, we create a knowledge space within the field of physics. By merging the node PACS code labels and the community tagging data, the results are depicted in Figure 1. \par

\begin{figure}[h]%
\centering
\includegraphics[width=0.9\textwidth]{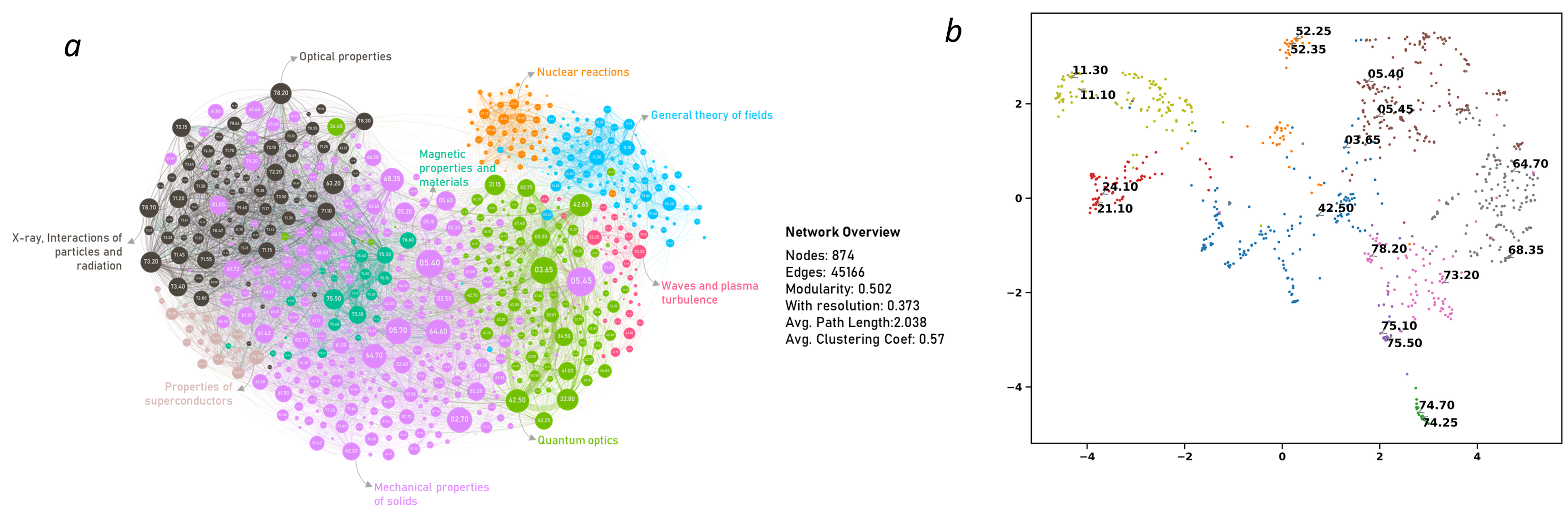}
\caption{The constructed knowledge space in Physics. a. The PACS code co-occurrence network. b. The embedded knowledge graph of PACS code co-occurrence network}\label{fig1}
\end{figure}

In Figure 1a, the physical subfields that share a community not only show remarkable proximity but also exhibit distinct clustering characteristics on the knowledge map. Each node in Figure 1a corresponds to a PACS code, where the node's size is determined by the number of connecting edges. The nodes are distinguished by different colors representing the identified 9 subfields. In this context, a higher co-occurrence frequency between PACS codes translates into a shorter distance in the network, thus indicating a closer knowledge relationship between those specific PACS codes. This is evident in the network as nodes belonging to the same community or a particular subfield are grouped closely together. \par
In addition, as shown in Figure 1b, the knowledge space effectively preserves the distinctions between different subfields. For example, the left side of the overall space is dominated by subfields related to condensed matter and statistical physics, and the right side is characterized by two subfields representing nuclear physics and astrophysics. It demonstrates that it is reasonable and effective to use the graph-embedded method to construct a knowledge map of physics.\par
After establishing the PACS code coordinates, we extract labeling information connecting authors' papers with PACS codes. Using this data, we calculate the center of mass for each paper, allowing us to position them on the knowledge map. 

\begin{figure}[h]%
\centering
\includegraphics[width=0.9\textwidth]{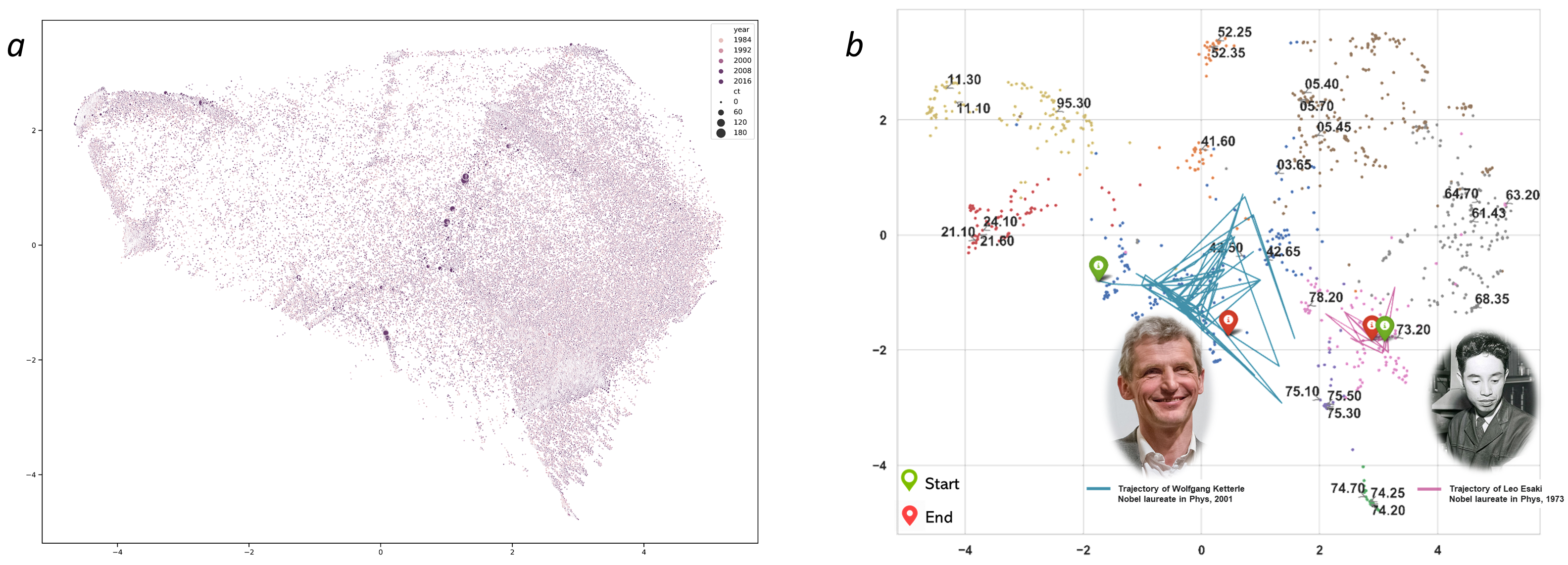}
\caption{The illustration of scientists' trajectories in the knowledge space. a. The distribution of papers in the physical knowledge field. b. Moving trajectories of two Nobel laureates\cite{wiki:ketterle2020,wiki:esaki2020}}\label{fig2}
\end{figure}

The distribution of papers in the physical field within the knowledge space is depicted in Figure 2. In the knowledge map of Figure 2a, scattered dots represent papers and colors indicating 9 subfields in physics. The topological structure of the field knowledge space, along with the location information of each paper on the map, serves as the foundational basis for quantitatively analyzing scientists' topic-transition. In Figure 2b, we illustrate the publication trajectories of two Nobel Prize laureates, Wolfgang Kettler (left, blue) and Leo Esaki (right, pink), within the physics field knowledge space. Wolfgang Kettler's Nobel Prize-winning contributions are in the realm of trapping cold atoms and reaching absolute zero, fundamental to the study of condensed matter within atomic physics. By observing his publication trajectory, we observe that his research encompasses nearly all subspaces of atomic physics. Leo Esaki's significant accomplishment lies in the discovery of the quantum tunneling effect in semiconductor materials, a key component of the superconductivity subfield in physics. In contrast to Wolfgang, Esaki's scientific exploration appears more focused on his research trajectory.\par
These findings underscore the divergent topic-transition trajectories of scientists within physics, despite their significant contributions to the field. This variation is likely attributed to the distinct research fields they inhabit. For the physics community as a whole, it remains fascinating to unravel the statistical patterns governing the selection and transition of research topics.
\subsection{The non-scale-free pattern of the aggregated inter-flow of scientists in the knowledge space}
When a researcher's paper transitions from one region of knowledge space to another, we can trace a sequence of origin and destination (OD) points within the region, mapping a trajectory from point \textit{i} to point \textit{j}. As we introduced before, we employ a partitioning of the knowledge space into two categories: the grid diagram and the voronoi diagram, following the spatial division principles of Geographic Information System (GIS) analysis.
\begin{figure}[h]%
\centering
\includegraphics[width=0.8\textwidth]{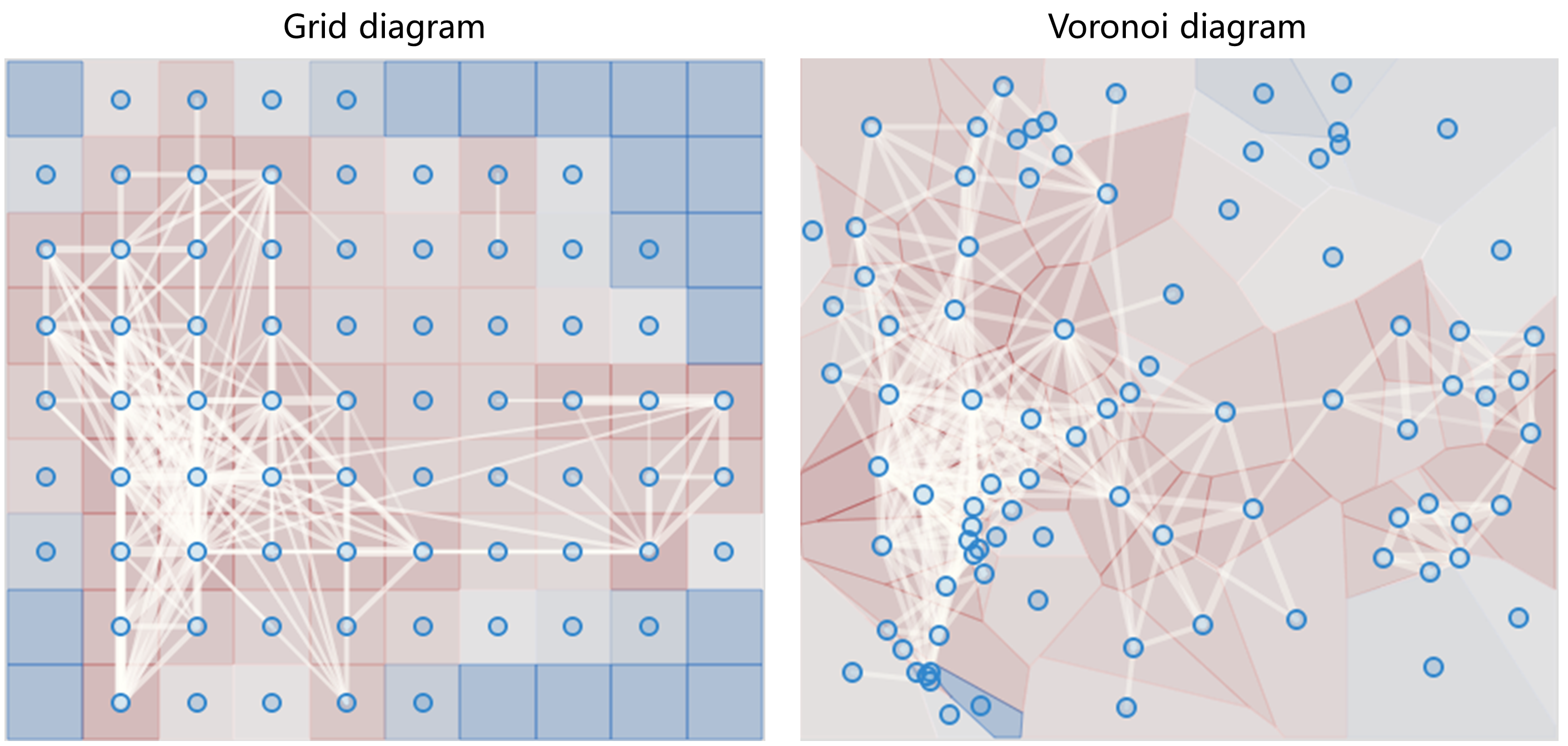}
\caption{The aggregated inter-flow of scientists in the knowledge space under two types of tessellations (inter-flow \(\geq\)150)}\label{fig3}
\end{figure}
Figure 3 illustrates these divisions: solid lines demarcate boundaries, circles signify central positions, while white connecting edges represent OD flows between regions, where the volume of flow is larger than 150. Furthermore, the color gradient of the subregion, transitioning from blue to red, indicates the incremental increase in the population size within each region. \par
\begin{figure}[h]%
\centering
\includegraphics[width=0.9\textwidth]{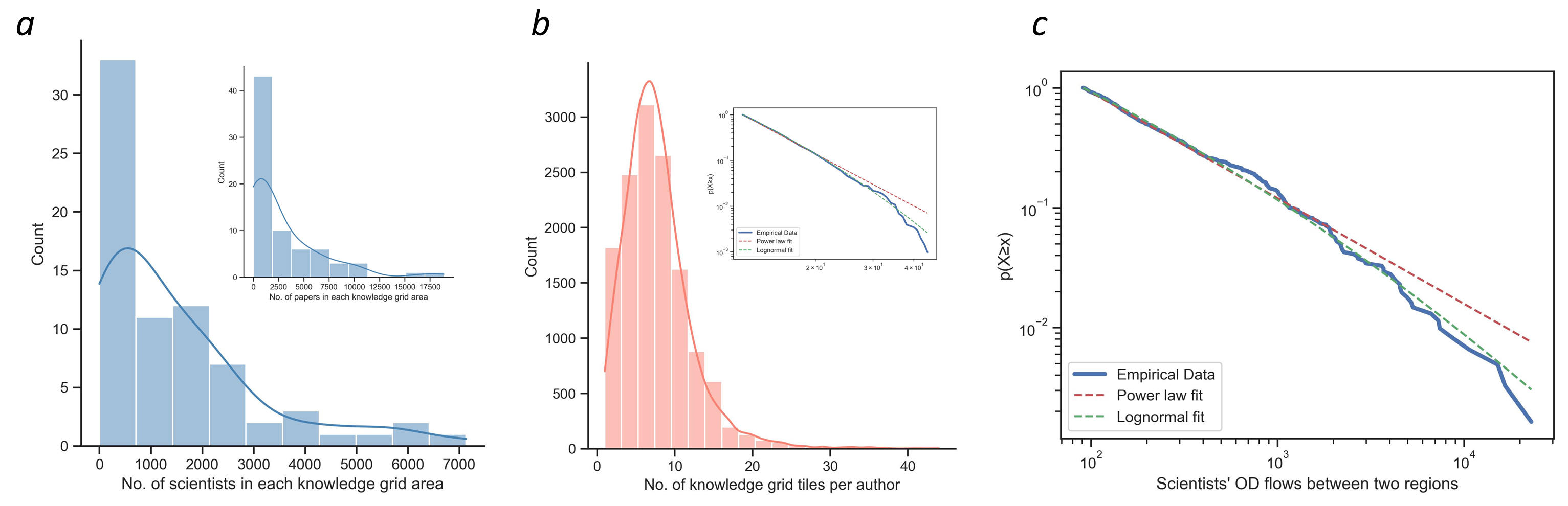}
\caption{Distribution of scientists' mobility characteristics in grid space
      a. Distribution of the number of scientists or papers in each grid area. b. A Log-norm distribution of the number of grid tiles for each scientist. c. A Log-norm distribution of OD flows from origin to destination}\label{fig4}
\end{figure}
In Figure 4, we present essential statistics on scientists' mobility within a grid space. It includes the distribution of the number of scientists or papers at each grid region, the number of scientists' knowledge tiles, and Origin-Destination (OD) flows. Moreover, using the power-law distribution fitting method proposed by Alstott et al.\cite{RN331}, our analysis reveals that the number of grid tiles associated with each scientist, and the corresponding origin-destination (OD) flow patterns, exhibit log-normal distributions rather than scale-free characteristics.
\begin{figure}[h]%
\centering
\includegraphics[width=0.9\textwidth]{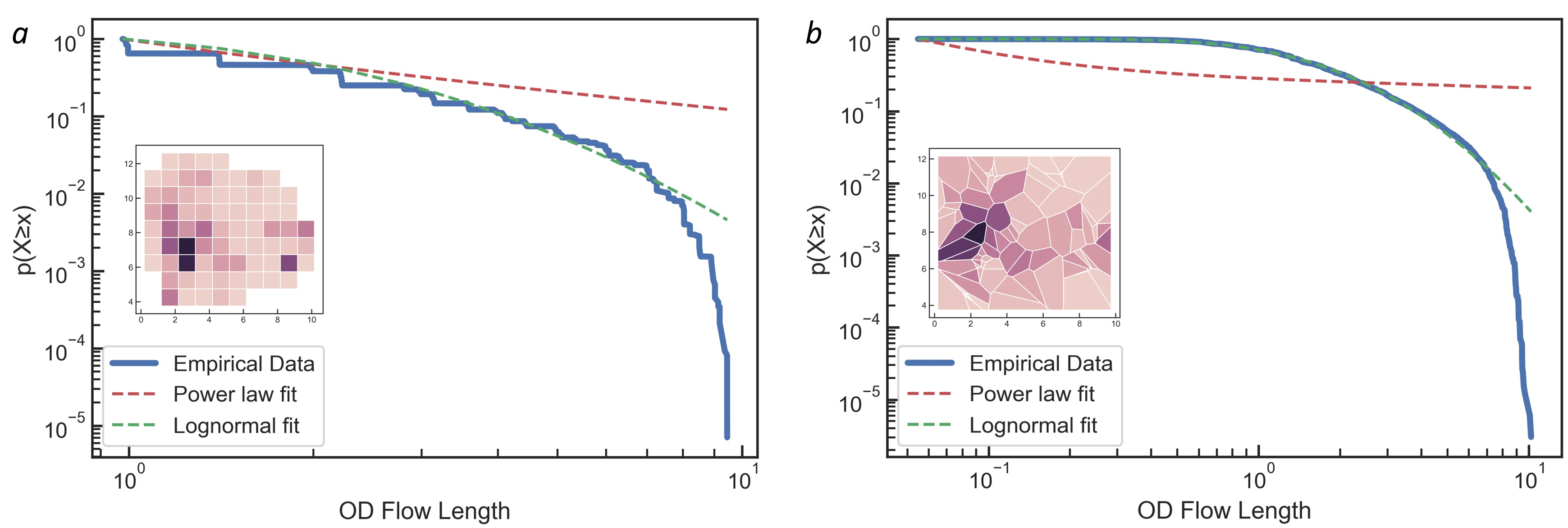}
\caption{The survival distribution function \textit{CCDF} of the OD distance of scientists' mobility in the knowledge space}\label{fig5}
\end{figure}
Figure 5 a-b depicts the distribution of OD flow distances originating from and ending at scientists' locations under grid and voronoi diagram partitioning methods. We also apply power-law and log-normal function fitting to the complementary cumulative characteristic distribution (\textit{CCDF}) of these OD flow distances. Furthermore, the insets in Figure 5 illustrate the density distribution of people within each spatial region.\par
Our analysis reveals that scientists' OD flow distance distribution exhibited more log-normal features than power-law characteristics under both the grid diagram and the voronoi diagram methods. Notably, the voronoi diagram partitioning method yields superior log-normal distribution fitting results compared to the power-law fit. This heavy-tailed distribution suggests that scientists' inter-field exploration patterns are not notably ‘scale-free’, despite being characterized by short-distance transitions for the majority and long-distance transitions for the minority.\par
\subsection{Models of Scientists' Topic-transition Behavioral Patterns}
Delving into the social factors that influence scientists' decisions to change their research topics is key to understanding the dynamics of scientific progress. To what extent can we predict scientists' topic-transition? Addressing this question requires a deep exploration of the behavioral mechanisms underlying group-level mobility patterns within the knowledge space. Building upon the established knowledge space and scientists' publication trajectories, we introduce two models within the framework of GIS analysis methodology: the gravity model and the radiation model. \par
\begin{figure}[h]%
\centering
\includegraphics[width=0.9\textwidth]{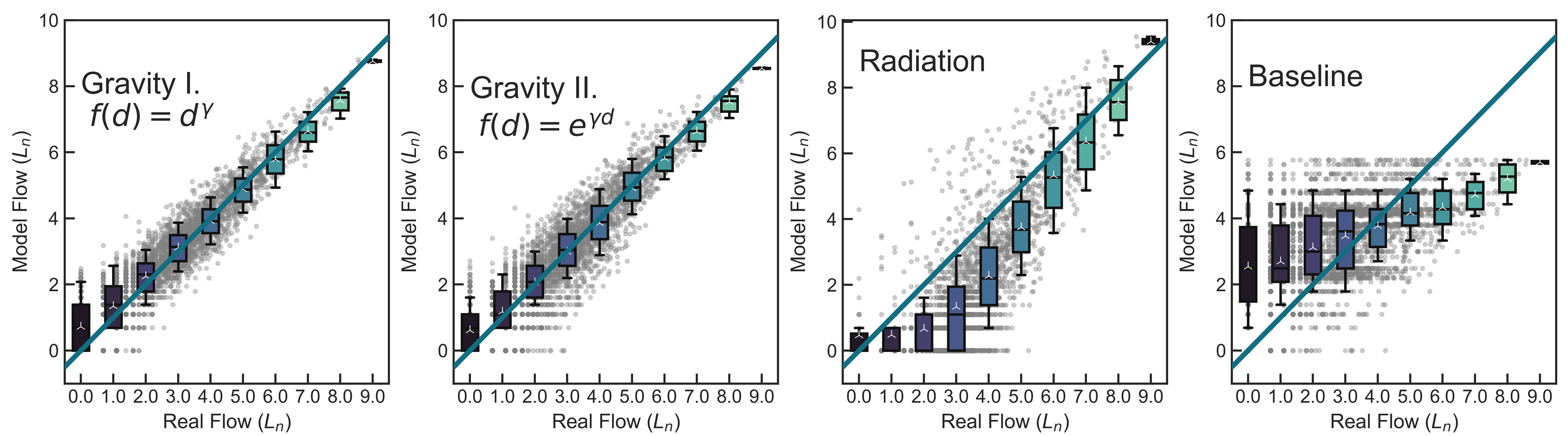}
\caption{The predicted OD flow results of scientists' topic transition models in the knowledge space}\label{fig6}
\end{figure}
Figure 6 presents a comparison between actual origin-destination (OD) flows and model-predicted flows across various types and parameters of population-level models. Gray points represent the correspondence level between observed and predicted flows for scientist topic-transition behaviors at each pair of starting and ending points. Box plots illustrate the 0.5-fold interquartile ranges, offering insights into data concentration intervals. White upward triangular symbols pinpoint the mean values of this dataset, and a green diagonal reference line represents a perfect alignment between actual and model results. The baseline model, where the damping function employs a \(\gamma\) parameter set to 0, effectively nullifying the impact of distance difference, performs the poorest in prediction accuracy. In contrast, both gravity models outperform the radiation model. The Box plot reveals that the exponential damping function in the gravity model yields superior predictions compared to the power-law damping function.\par
\begin{figure}[h]%
\centering
\includegraphics[width=0.9\textwidth]{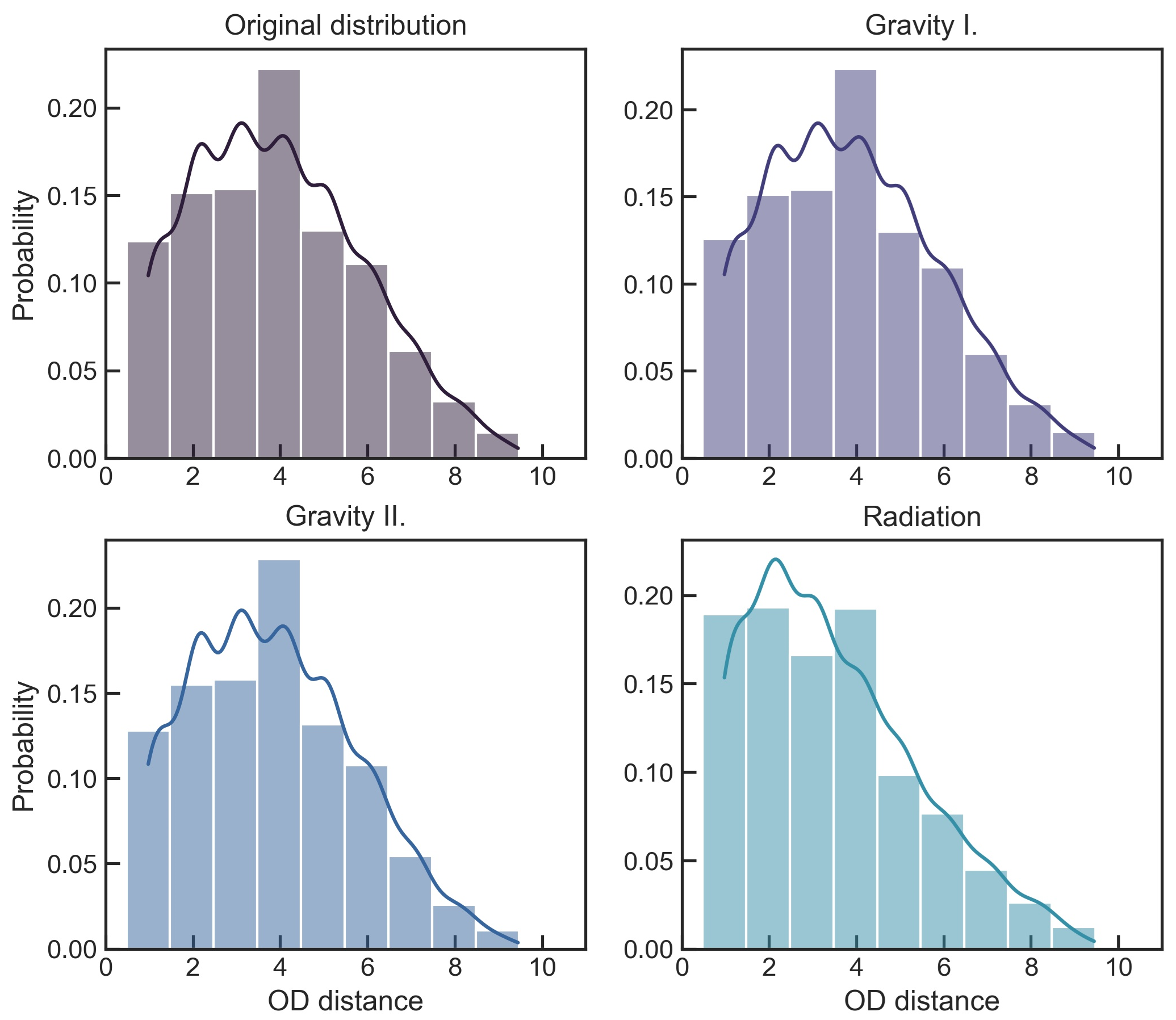}
\caption{The predicted OD distance distribution of three collective-level scientists' topic transition models}\label{fig7}
\end{figure}
Figure 7 displays the observed OD distance density distributions in knowledge space alongside three model-predicted distributions. Our analysis reveals that the gravity model again offers a superior capability of explanations and predictions for the patterns of scientists' topic-transition within the knowledge space, compared to the radiation model. To ensure the robustness and consistency of our findings, we conduct experiments involving adjustments to the division scale of the field knowledge space and introduce randomized experiments in various contexts. These results serve to scrutinize the model predictions further.\par
\begin{figure}[h]%
\centering
\includegraphics[width=0.9\textwidth]{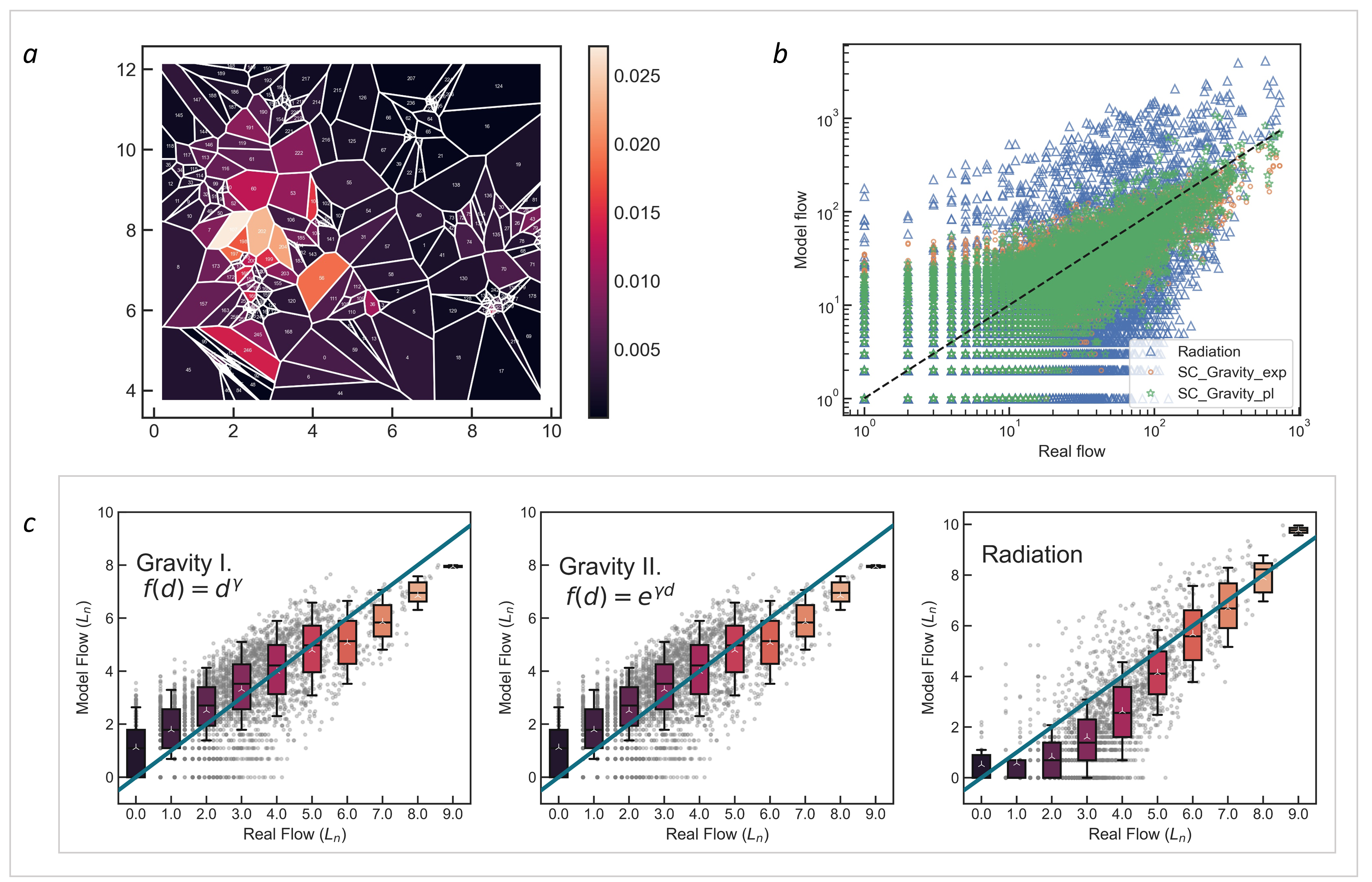}
\caption{The results of robustness experiments of scientists' topic transition model in the knowledge space. a-b. a fine-grained voronoi diagram of knowledge space with 258 subspaces and its predicting results of topic-transition models. c.The predicting results of topic-transition models under the experiment of randomizing papers' coordinates}\label{fig8}
\end{figure}
In our scale-reconfiguration experiments (see Figure 8), we alter the scale of subfield regions by different multiples, and subsequently reevaluate the topic-transition pattern of scientists as well as the predictions from the simulation model. Figure 8a illustrates the subdivision of the voronoi diagram into smaller segments, expanding the high-frequency 10 PACS codes from each subfield community to 30 PACS codes, creating 258 non-empty subspaces. Figure 8b compares actual OD flows with model predictions at this scale setting, showing the continued superiority of gravity models over the radiation model.  \par
In the null model experiments, three scenarios were tested: 1) randomizing authors’ publication date order to remove sequential timing effects, 2) random perturbation of paper coordinates in the knowledge space, and 3) maintaining the author's publication frequency while randomly selecting the same number of papers. Figure 8c demonstrates the diminished results of scenario 2 in the randomized experiment, highlighting the significance of keeping original publication coordinates in the knowledge space for predicting OD flows. The simulation results of scenarios 1 and 3 are not shown but close to scenario 2.\par
It's important to note that the key distinction between the gravity model and the radiation model lies in key factors that drive scientists’ mobility in the knowledge space. The gravity model emphasizes the impact of distance between subfield regions on topical transition, while the radiation model focuses on attraction or repulsion based on potential research gaps between subfield regions. Our findings suggest that the distance between subfields and the number of scientists in research subfields have a more significant influence on scientists' movement than the potential research ‘opportunities’ between subfields. Although peripheral research areas between subfield regions are crucial for scientific progress but pose risks, as their outcome is unpredictable. This uncertainty may contribute to the radiation model's reduced predictive accuracy, while the gravity model aligns with most scientists' conservative and ‘hot-spot-tracing’ research strategy when selecting or transiting research topics.\par
\subsection{Null model experiments and Robustness test of results}
We systematically assess the effect of different parameters or experimental settings on model performance, including subfield region division granularity, damping function types, and randomized permutations in authors' trajectories. In addition, we introduce multiple model evaluation indices to compare experimental results comprehensively.\par

\begin{table}[h!]
\setlength\tabcolsep{1.8 pt}
\caption{The aggregated results of the evaluation indexes of two population-level models and null models}
      \begin{tabular}{cccccccccccc}
      \toprule
Id & Network 
& Robust & Model & Model& \(R^2\) & RMSE & Spearman& Pearson & CPC & CPCd & CPL\\
& Type &  Exp. & & Para. & & &  Coef. & Coef. \\

\midrule
1 & Grid & BSL & Baseline & - & 0.088 & 269.57 & 0.475 & 0.324 & 0.391 & 0.006 & 1\\
2 & Grid & BSL & Gravity & exp & 0.888 & 94.625 & 0.906 & 0.944 & 0.8 & 0.011 & 1\\
3 & Grid & BSL & Gravity & pl & 0.887 & 94.778 & 0.935 & 0.943 & 0.82 & 0.011 & 1\\
4 & Grid & BSL & Radiation & - & -0.541 & 350.498 & 0.843 & 0.729 & 0.534 & 0.007 & 0.755\\
5 & Grid & Rand1 & Gravity & exp & 0.878 & 98.478 & 0.913 & 0.94 & 0.794 & 0.011 & 1\\
6 & Grid & Rand1 & Gravity & pl & 0.892 & 92.805 & 0.932 & 0.945 & 0.818 & 0.011 & 1\\
7 & Grid & Rand2 & Gravity & exp & 0.527 & 1914 & 0.761 & 0.73 & 0.572 & 0.011 & 1\\
8 & Grid & Rand2 & Gravity & pl & 0.527 & 191 & 0.761 & 0.731 & 0.572 & 0.011 & 1\\
9 & Grid & Rand3 & Gravity & exp & 0.529 & 193.76 & 0.762 & 0.732 & 0.573 & 0.011 & 1\\
10 & Grid & Rand3 & Gravity & pl & 0.53 & 193.635 & 0.762 & 0.733 & 0.573 & 0.011 & 1\\
11 & Grid & Scale1 & Gravity & exp & 0.828 & 22.865 & 0.775 & 0.919 & 0.76 & 0.065 & 1\\
12 & Grid & Scale1 & Gravity & pl & 0.866 & 20.201 & 0.797 & 0.931 & 0.785 & 0.069 & 1\\
13 & Grid & Scale1 & Radiation & - & -2.284 & 100.007 & 0.645 & 0.67 & 0.431 & 0.039 & 0.494\\
14 & Grid & Scale2 & Gravity & exp & 0.878 & 502.72 & 0.954 & 0.938 & 0.839 & 0.001 & 1\\
15 & Grid & Scale2 & Gravity & pl & 0.851 & 555.99 & 0.968 & 0.923 & 0.842 & 0.001 & 1\\
16 & Grid & Scale2 & Radiation & - & 0.18 & 1304.34 & 0.92 & 0.713 & 0.603 & 0.001 & 0.923\\
17 & voronoi & BSL & Baseline & - & 0.092 & 128.097 & 0.421 & 0.304 & 0.424 & 0.008 & 1\\
18 & voronoi & BSL & Gravity & exp & 0.836 & 54.514 & 0.857 & 0.917 & 0.79 & 0.017 & 1\\
19 & voronoi & BSL & Gravity & pl & 0.769 & 64.61 & 0.866 & 0.879 & 0.77 & 0.018 & 1\\
20 & voronoi & BSL & Radiation & - & -1.685 & 220.29 & 0.827 & 0.679 & 0.488 & 0.011 & 0.761\\
21 & voronoi & Rand1 & Gravity & exp & 0.821 & 56.869 & 0.863 & 0.911 & 0.784 & 0.018 & 1\\
22 & voronoi & Rand1 & Gravity & pl & 0.781 & 62.963 & 0.866 & 0.884 & 0.768 & 0.018 & 1\\
23 & voronoi & Rand2 & Gravity & exp & 0.4 & 104.114 & 0.648 & 0.642 & 0.539 & 0.015 & 1\\
24 & voronoi & Rand2 & Gravity & pl & 0.401 & 104.039 & 0.648 & 0.642 & 0.539 & 0.015 & 1\\
25 & voronoi & Rand3 & Gravity & exp & 0.401 & 103.997 & 0.649 & 0.643 & 0.539 & 0.015 & 1\\
26 & voronoi & Rand3 & Gravity & pl & 0.403 & 103.886 & 0.649 & 0.644 & 0.54 & 0.015 & 1\\
27 & voronoi & Scale1 & Gravity & exp & 0.745 & 14.502 & 0.766 & 0.874 & 0.746 & 0.08 & 1\\
28 & voronoi & Scale1 & Gravity & pl & 0.708 & 15.523 & 0.76 & 0.845 & 0.748 & 0.088 & 1\\
29 & voronoi & Scale1 & Radiation & - & -5.169 & 71.326 & 0.669 & 0.581 & 0.38 & 0.046 & 0.476\\
30 & voronoi & Scale2 & Gravity & exp & 0.836 & 5187 & 0.857 & 0.917 & 0.79 & 0.017 & 1\\
31 & voronoi & Scale2 & Gravity & pl & 0.776 & 63.388 & 0.868 & 0.883 & 0.771 & 0.019 & 1\\
32 & voronoi & Scale2 & Radiation & - & -1.702 & 220.143 & 0.829 & 0.682 & 0.488 & 0.011 & 0.762\\
33 & voronoi & Scale3 & Gravity & exp & 0.836 & 54.349 & 0.857 & 0.917 & 0.79 & 0.017 & 1\\
34 & voronoi & Scale3 & Gravity & pl & 0.769 & 64.412 & 0.866 & 0.879 & 0.769 & 0.018 & 1\\
35 & voronoi & Scale3 & Radiation & - & -1.701 & 220.249 & 0.827 & 0.681 & 0.488 & 0.011 & 0.762\\
        \bottomrule
      \end{tabular}

\begin{tabular}{l}
\textbf{Note}: 1. Spearman and Pearson coefficients in the experiment p-value are less than 0.001. \\2. Abbreviations of BSL: benchmark experiment under the initial setting, exp: exponential function, \\pl: power rate function, Rand: randomized experiment, Scale: scale expansion/reduction experiment. 
\end{tabular}
\end{table}

As summarized in Table 1, we deploy experiments with specific groups to evaluate model predictions against real results under various experimental conditions. Experiment groups 1-4 and 17-20 correspond to basic experimental settings depicted in Figures 5-8. Experiment groups 5-10 and 21-26 involve randomized experiments with grid-based diagram and voronoi-based diagram division, respectively, aligning with the above null model experiments. Experiment groups 11-16 explore model evaluation with grid region granularity reduced and expanded by a factor of 1. Experiment groups 27-32 pertain to modeling the voronoi diagram subregions, involving adjustments to the number of high-frequency PACS codes and corresponding sub-regions. Furthermore, we consider the impact of coordinate scale transformations on experimental predictions, with experiments 33-35 representing scaled experiments.\par
Cross-validating across different model evaluation metrics minimizes bias inherent to a single metric. Of particular interest is the CPC indicator, widely used in the studies of human mobility behavior at the collective level, measuring explorer's overlap trajectories between origins and destinations in real or model-predicted data. By comparing various model evaluation metrics in Table 4.1, we deduce five key findings:\par
(1) Regardless of the grid partition type and subregion granularity, two gravity models significantly outperform the radiation model, predicting over 30\% more real OD flows and 25\% more trajectories. The baseline model, which does not consider distance factors, produces the poorest predictive results, with CPC indices of only 0.391 and 0.424 in the grid and voronoi diagram cases, respectively.\par
(2) In the scale experiments, while the predictive power of the gravity model decreases with a smaller unit area granularity and increases with a larger granularity, overall, the scaling of the model does not significantly impact predictive performance. The minimum CPC index remains around 0.75.\par
(3) Regarding the three sets of null model experiments, only the model generated by shuffling the order of authors' publications shows a slight decrease in predictive performance compared to the baseline model, with a decrease of only 0.01 in the CPC index. However, the two models created by randomly shuffling all paper coordinates exhibit a noticeable drop in predictive performance for real OD flows, with a reduction of 0.23 in the CPC index.\par
(4) When uniformly reducing the coordinate scale by a factor of 10 without changing the grid partition granularity, the predictive power of the model remains largely unchanged. The experimental results of groups 33-35 show only minor differences compared to groups 18-20.\par
(5) In terms of the damping function type in the gravity model, the exponential function model under the grid partition is slightly inferior to the power-law function model in predicting results, whereas the results are reversed under the voronoi diagram partition. \par
\textbf{\begin{figure}[h]%
\centering
\includegraphics[width=0.9\textwidth]{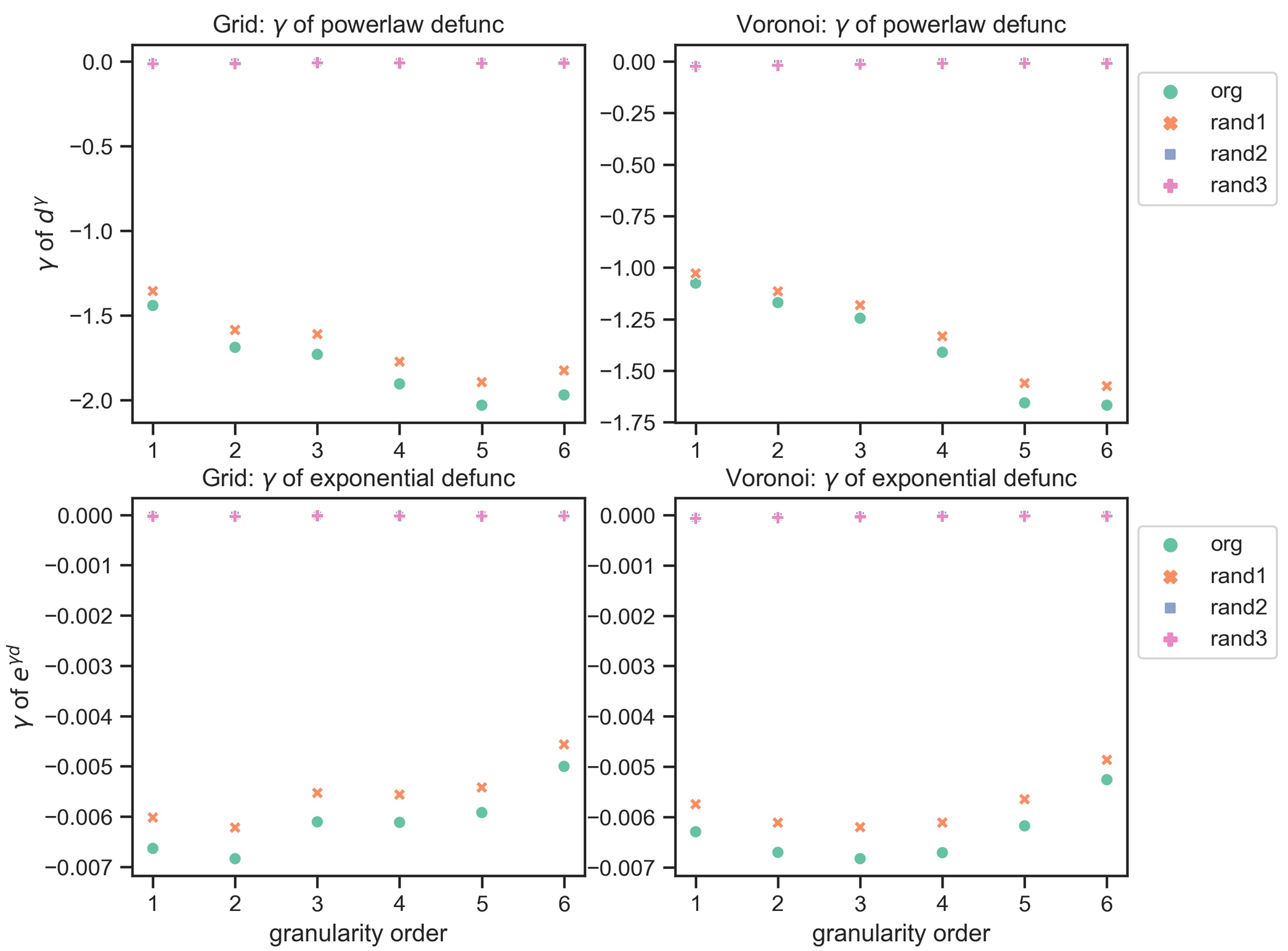}
\caption{The analysis of the distance exponent \(\gamma\) in the deterrence function under the different randomly configured models}\label{fig9}
\end{figure}}
Furthermore, we analyze the relationship between different levels of granularity in knowledge space partitioning, including three different random experiments, and the \(\gamma\) index in the gravity model damping function. As shown in Figure 9, the analysis reveals that in the context of real scientists' topic selection and transition within the knowledge space, the absolute value of the distance decay factor \(\gamma\) between scientists in different regions exceeds that in three other random experimental scenarios. This result underscores a significant bounded characteristic in the transition of scientists' interests. The conserved characteristic is influenced by mixed factors such as modularized knowledge structure, individual knowledge attributes, exploration preference patterns, or inter-domain knowledge barriers as scientists move in the knowledge space.\par
\begin{figure}[h]%
\centering
\includegraphics[width=0.99\textwidth]{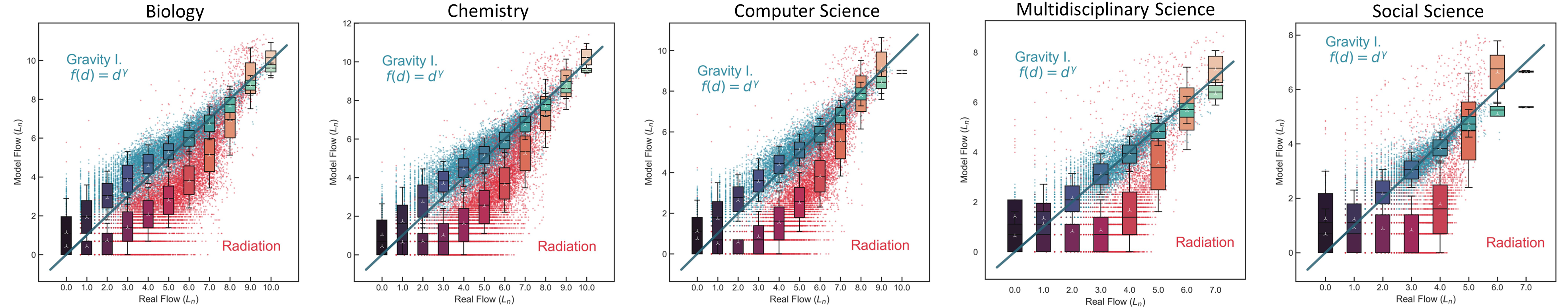}
\caption{The predicted results of scientists’ topic transition model in the disciplines of Biology, Chemistry, Computer Science, Multidisciplinary Science, and Social Science}\label{fig10}
\end{figure}
To assess the generalizability of our findings beyond physics, we test the performance of the gravity model and the radiation model across diverse disciplines. As depicted in Figure 10, the results demonstrate the robustness of our proposed gravity model compared to the radiation model across various fields, including Biology, Chemistry, Computer Science, Multidisciplinary Science, and Social Science. In all disciplines, the gravity model consistently outperforms the radiation model in predicting scientists' actual mobility patterns within the knowledge space.\par
However, further examination of the simulation results depicted in the grid diagram reveals a significant variance in model performance across disciplines. Social Science exhibits the lowest R-squared metric of 0.746 (p\(<\).001), while Chemistry achieves the highest R-squared metric of 0.874 (p\(<\).001). The observed disciplinary discrepancies reveal diverse patterns in scientists' exploration paths within the knowledge space.\par
\section{Discussion}
In this study, we utilize a knowledge space to map the trajectories of scientists' publications in chronological order, shedding light on their patterns of topic selection and transition within this knowledge space. We subdivide this space into grid or voronoi diagram subfields using density and equidistant approaches. Our analysis reveals an overall log-normal distribution of scientists' topic-transition distances at the origins and destinations. To delve into the mechanisms governing these topic transitions at a group level, we introduce two movement behavior models: the gravity and radiation models. Our findings indicate that the gravity model, driven by factors such as population size and knowledge distance, outperforms considerations of research gap areas in explaining and predicting scientists' topic-transition behaviors. To enhance our insights, we compare our results to three key aspects related to existing studies: \par
1. Comparison with human commuting patterns in real geographic space: We find that scientists' explorations in the knowledge space are more influenced by ‘distance’ and regional ‘population’ factors than ‘opportunity’ factors. This mirrors the patterns observed in human commuting within administrative regions in a city, albeit without predefined sub-field spaces in our knowledge space.\par
2. Comparison with human movement patterns in virtual space: Scientists' exploratory behavior in the knowledge space exhibits similarities to human behaviors in virtual spaces. The log-normal distribution of exploration trajectories aligns with patterns seen in the game and website access behaviors\cite{RN1438}. Although the space construction frameworks differ, the underlying psychological mechanisms for resource search and acquisition appear to share commonalities\cite{Wang2017}.\par
3. Comparison with other models of scientists' topic-changing or switching behavior: We emphasize a collective rather than individual-level perspective on scientists' topic selection and transition, and find that knowledge distance and population size are two key social factors in explaining scientists' exploration patterns in the knowledge space, suggesting a typical hotspot-tracing tendency for the majority of scientists.\par
In summary, our research advances the understanding of scientists' topic transition by accounting for social influences and distance heterogeneity in the constructed knowledge space. Our findings suggest that most scientists tend to make cautious topic transitions, guided primarily by the number of scientists in their field and the knowledge distance between fields, rather than by ‘gaps’ or ‘opportunities’. This cautious approach may have significant implications for the efficiency and effectiveness of the scientific innovation system. \par
\section{Conclusion}
Our study deploys quantitative analysis methods to investigate scientists' topic selection and transitions, offering insights into the underlying mechanisms at the group level. We find that scientists' movements within the knowledge space exhibit heterogeneity, characterized by an overall log-normal distribution of OD flow distances. It indicates that, in essence, most scientists tend to make prudent and short-range transitions in their research interests. Our analysis identifies key social factors, including subfield population size, research gaps or opportunities, and knowledge distances, as instrumental in shaping scientists' topic transition.\par
The mechanistic analysis reveals a prevailing tendency towards hotspot-tracing and opportunity-seeking within the academic field, akin to animal foraging behavior, where resource distribution influences foraging strategies. In the competitive realm of scientific research, adopting a conservative strategy appears safe for scientists. Most scientists tend to follow a hotspot-tracing tendency rather than proactively exploring research opportunities between subfields and connecting knowledge from different domains. This conservatism can lead to issues like resource concentration, reduced research originality, and decreased research efficiency for the whole scientific enterprise. Understanding this conservative strategy reveals valuable insights into the dynamics of scientists' knowledge-creation within the innovation system, and provides empirical support for science policymakers. \par
In future research, we plan to refine existing population-level models by incorporating additional factors that influence scientific mobility, such as individual career aspirations, hotspots’ knowledge structures, and the evolving landscape of scientific research, optimize model performance by exploring various machine learning algorithms, and investigate the nuances of scientific mobility across diverse disciplines and career stages, utilizing academic datasets spanning a broad range of fields and historical periods.

\bigbreak
\backmatter
\bmhead*{Acknowledgements}
Not applicable.

\bmhead*{Funding}
This work is supported by the National Natural Science Foundation of China under Grant Nos. 72371052 and 71871042 (to HX), and by the Humanities and Social Science Project of the Ministry of Education of China Grant No 18YJA630118 (to HX). 

\bmhead*{Availability of data and materials}
The APS data are available at https://journals.aps.org/datasets by submitting a request. The MAG data used in this paper was downloaded via the Microsoft Academic Graph APIs. However, the Microsoft Academic website and underlying APIs have been retired in 2021. All other materials used in this study are available from the corresponding author upon reasonable request.

\bmhead*{Competing interests}
  The authors declare that they have no competing interests.

\bmhead*{Author's contributions}
HX and FL conceived the study. FL and SZ designed the research. FL and SZ performed the experiments. All authors contributed to the analysis of the results and writing of the manuscript.

\bibliography{sn-article}
\end{document}